 \documentclass[final,5p,times,twocolumn,authoryear]{elsarticle}

\usepackage{amssymb}
\usepackage{lipsum}
\usepackage{hyperref}
\usepackage{xcolor}

\newcommand{\Gaia}{\textit{Gaia} }

\journal{New Astronomy}

\begin{document}

\begin{frontmatter}

%% Title, authors and addresses

%% use the tnoteref command within \title for footnotes;
%% use the tnotetext command for theassociated footnote;
%% use the fnref command within \author or \affiliation for footnotes;
%% use the fntext command for theassociated footnote;
%% use the corref command within \author for corresponding author footnotes;
%% use the cortext command for theassociated footnote;
%% use the ead command for the email address,
%% and the form \ead[url] for the home page:
%% \title{Title\tnoteref{label1}}
%% \tnotetext[label1]{}
%% \author{Name\corref{cor1}\fnref{label2}}
%% \ead{email address}
%% \ead[url]{home page}
%% \fntext[label2]{}
%% \cortext[cor1]{}
%% \affiliation{organization={},
%%            addressline={}, 
%%            city={},
%%            postcode={}, 
%%            state={},
%%            country={}}
%% \fntext[label3]{}

\title{How \textit{Gaia} sheds light on the Milky Way star cluster population}

%% use optional labels to link authors explicitly to addresses:
%% \author[label1,label2]{}
%% \affiliation[label1]{organization={},
%%             addressline={},
%%             city={},
%%             postcode={},
%%             state={},
%%             country={}}
%%
%% \affiliation[label2]{organization={},
%%             addressline={},
%%             city={},
%%             postcode={},
%%             state={},
%%             country={}}

\author[first]{T. Cantat-Gaudin}
\affiliation[first]{organization={Max Planck Institute for Astronomy},%Department and Organization
            addressline={Königstuhl 17}, 
            postcode={69 117}, 
            city={Heidelberg},
            country={Germany}}

\author[second]{L. Casamiquela}
\affiliation[second]{organization={GEPI, Observatoire de Paris, PSL Research University, CNRS, Sorbonne Paris Cité},%Department and Organization
            addressline={5 place Jules Janssen}, 
            postcode={92190}, 
            city={Meudon},
            country={France}}

\begin{abstract}
%% Text of abstract
Star clusters are among the first celestial objects catalogued by early astronomers. As simple and coeval populations, their study has been instrumental in charting the properties of the Milky Way and providing insight into stellar evolution through the 20th century. Clusters were traditionally spotted as local stellar overdensities in the plane of the sky. In recent decades, for a limited number of nearby clusters, it became possible to identify cluster members through their clustering in proper motion space. With its astrometric data of unprecedented precision, the \textit{Gaia} mission has completely revolutionised our ability to discover and characterise Milky Way star clusters, to map their large-scale distribution, and to investigate their internal structure. In this review we focus on the population of open clusters, residing in the Galactic disc. We summarise the current state of the \textit{Gaia}-updated cluster census and studies of young clusters and associations. We discuss recent developments in techniques for cluster detection and age estimation. We also review results enabled by \textit{Gaia} data concerning the dynamical evolution of gravitationally bound clusters and their stellar inventory.
\end{abstract}

%%Graphical abstract
%\begin{graphicalabstract}
%\includegraphics{grabs}
%\end{graphicalabstract}

%%Research highlights
%\begin{highlights}
%\item Research highlight 1
%\item Research highlight 2
%\end{highlights}

\begin{keyword}
%% keywords here, in the form: keyword \sep keyword, up to a maximum of 6 keywords
Milky Way \sep open clusters \sep data mining 

%% PACS codes here, in the form: \PACS code \sep code

%% MSC codes here, in the form: \MSC code \sep code
%% or \MSC[2008] code \sep code (2000 is the default)

\end{keyword}

\end{frontmatter}

%\tableofcontents

%% \linenumbers

\section{Introduction} \label{sec:introduction}

Historically, stellar clusters have been identified as local overdensities of stars in a given region of the sky. 
It has long been recognised that young stars tend to be found near other young stars \citep{Bok34,Blaauw52,Blaauw64}, which led to the common (and perhaps oversimplified) assumption that all stars were, at least in their infancy, aggregated with siblings who formed from the collapse of the same parent molecular cloud \citep{Lada91,Lada03, Kruijssen11,Pfalzner12, Parmentier13,Kamdar19born}. When these groups are formed sufficiently dense \citep[$\sim$1\% to 70\% of the clusters in spiral galaxies depending on the local gas density, according to][]{Kruijssen12} they can remain gravitationally bound for hundreds of millions of years. In the Milky Way, it is estimated that $\sim$4\% to 14\% of the total stellar mass comes from once-gravitationally-bound clusters \citep{Goddard10}, and 30-35\% in the entire Universe \citep[][]{Kruijssen12}.

Star clusters sit at a crossroads of scales, and are relevant to many aspects of astronomy and astrophysics. As coeval and chemically homogeneous groups, they are routinely used as tracers of Galactic structure and evolution, and as calibrators for the distance scale and for developing stellar
evolution models. In this review we focus on objects traditionally referred to as open clusters. They are less dense and less massive than globular clusters (a few 100 to a few 1000 solar masses), younger (a few Myr to several Gyr), and follow Galactic orbits typical of the $\alpha$-poor, metal-rich disc from which they originate.

The ESA \textit{Gaia} space mission \citep{Perryman01,Gaia16prusti} and its first \citep[][]{GaiaDR1}, second \citep[DR2;][]{GDR2content} and third \citep{GaiaEDR3,GaiaDR3} data releases have had a huge impact on our knowledge of the cluster population in the Milky Way. The \Gaia catalogues have enabled the detection and discovery of a large number of new objects, and allowed us to characterise them with unprecedented precision. 

In this article we review the major advances brought by \Gaia in our understanding of Milky Way star clusters, and how they in turn impact our understanding of the structure and evolution of our Galaxy. 
Since 2018, over a hundred scientific papers exploiting \Gaia data have been published every month. This deluge of results makes it unfeasible for a such a review to be exhaustive, even restricting its scope to the topic of Galactic star clusters. We refer the reader interested in adjacent topics to the reviews of \cite{Wright20} and \citet{2022arXiv220310007W} on OB associations, \citet{2022Univ....8..111C} summarising \Gaia and star clusters until 2021, and the recent publication of \citet{Zucker2023ASPC..534...43Z} about the Solar neighbourhood in the \Gaia era.

This review discusses updates to the cluster census in Sect.~\ref{sec:census}. Section~\ref{sec:young} is dedicated to young clusters and associations. Section~\ref{sec:ages} presents various methods used to estimate the age of clusters. In Sect.~\ref{sec:dynamics} we review recent publications concerning the dynamical evolution of clusters.  Finally, Sect.~\ref{sec:spectro} discusses stellar evolution and chemical abundances through the lens of cluster studies.

\section{The cluster census} \label{sec:census}

\subsection{Identifying clusters and their members}

About a third of the open clusters known before \Gaia were catalogued by Charles Messier \citep{1781cote.rept..227M}, William Herschel \citep{1786RSPT...76..457H,1789RSPT...79..212H,1802RSPT...92..477H} and John Herschel \citep{1864RSPT..154....1H}, before being compiled into the \textit{New General Catalogue} \citep[NGC;][]{1888MmRAS..49....1D}. New observational techniques such as the advent of photographic plates did not have a major impact on the star cluster census, but serendipitous discoveries were added to cluster catalogues through the 20$^\textrm{th}$ century \citep[e.g.][]{1930LicOB..14..154T,1931AnLun...2....1C,1958csca.book.....A, 1970csca.book.....A, 1982A&A...109..213L,1985IAUS..106..143L}.
More recently, it became possible to discover new clusters as co-moving groups of stars in proper-motion catalogues \citep[e.g.][]{Alessi03,Kharchenko2005A&A...440..403K,Froebrich07}. Before the second \Gaia data release (in April 2018), the two most widely used catalogues of open clusters were \citet{Dias02} listing about 2000 objects, and \citet{Kharchenko13} listing over 3000, although significant fraction were stellar overdensities awaiting confirmation from more precise astrometric surveys.

The first \Gaia data release and its Tycho-\Gaia Astrometric Solution \citep[TGAS;][]{Michalik15} provided improved proper motions and, for the first time, parallaxes, for the two million stars of the Tycho-2 catalogue \citep{2000A&A...355L..27H}. This astrometric data set was however limited to magnitude $\sim$12, and the precision of its proper motions was only greater than the ground-based proper motion catalogue UCAC4 \citep{Zacharias2013AJ....145...44Z} in some regions of the sky privileged by the \Gaia scanning law, only allowing for the astrometric characterisation of a hundred clusters within 1\,kpc \citep{CantatGaudin18tgas}, and a detailed study of the Hyades \citep[46\,pc from us;][]{Reino2018MNRAS.477.3197R}

The second \Gaia data release \citet{GDR2content} was an immediate game changer in virtually all aspects of Milky Way studies, providing proper motions better than existing catalogues by a factor of 100, and a full astrometric solution for over 1 billion stars down to magnitude $\sim$20. Attempting to identify all $\sim$3000 known clusters in the Milky Way disc, \citet{CantatGaudin18gdr2} were only able to detect 1169 of them, indicating that a large number of clusters listed in the literature are putative groupings with no physical reality. The existence of many of them had in fact already been questioned \citep[by e.g.][when building the Revised New General Catalogue]{1973rngc.book.....S} and sometimes even convincingly refuted \citep[e.g. four NGC objects by][on the basis of incoherent radial velocities]{2018MNRAS.480.5242K}. \citet{CantatGaudin20mirages} have shown that many of these asterisms are created by extinction patterns in the inner regions of the Milky Way, leading to a fictitious population of old, inner-disc, high-altitude clusters in the Galactic disc. The presence of this old, inner-disc population had been difficult to explain theoretically \citep{2016ApJ...817L...3M} and made the Milky Way appear too rich in old clusters compared to other spiral Galaxies. Based on the \textit{Gaia}-updated cluster census, \citet{Anders21} have shown that the cluster-age function of the Milky Way is in fact in line with empirical expectations.

The unprecedented precision of the \Gaia data have allowed for serendipitous discoveries of new clusters \citep{CantatGaudin18gdr2,Ferreira19, Jaehnig21,Negueruela21} or discoveries through visual inspection of proper-motion diagrams \cite{Sim19}. 
Since star clusters are expected to be compact\footnote{In Sect.~\ref{sec:dynamics} we mention techniques used to recover structures when they are coherent but not compact.} on the sky ($\alpha$,$\delta$), in proper motion ($\mu_{\alpha*},\mu_{\delta}$) and in parallax ($\varpi$), many off-the-shelf clustering algorithms that were not initially designed for astronomical data can be applied directly to the \Gaia catalogue. Some examples include Density-Based Spatial Clustering of Applications with Noise \citep[DBSCAN;][]{DBSCANester1996density}, Hierarchical-DBSCAN \citet{HDBSCAN}, or Ordering Points To Identify the Clustering Structure \citep[OPTICS;][]{AnkEtAl99}. The most successful searches were however performed with carefully chosen data mining schemes and algorithms to pick up clusters in the large 5D \Gaia catalogue, based for instance on DBSCAN \citep[e.g.][]{Hunt21, CastroGinard18, CastroGinard19, CastroGinard20,2022A&A...661A.118C}, Gaussian Mixture Models \citep{CantatGaudin19coin}, or HDBSCAN \citep{2023A&A...673A.114H}. Over 4000 objects with high probability of being true clusters (according to their astrometric and photometric properties) are currently listed in the catalogue of \citet{2023A&A...673A.114H}. Several thousand more potential clusters have been identified in dozens of studies\footnote{We refer the reader to Table~3 of \citet{2023A&A...673A.114H} and Table~1 of \citet{2023MNRAS.526.4107P}  for lists of papers reporting cluster candidates.}, and might be confirmed as true clusters by the improved astrometry of the fourth \Gaia data release.

The nominal precision of the \Gaia proper motions can vary from $\sim$0.01\,mas\,yr$^{-1}$ for bright and well-behaved sources to over 0.5\,mas\,yr$^{-1}$. Clusters can be found at distances ranging from a few parsecs to 15\,kpc from us, and exhibit a variety of sizes, densities, and surrounding environment. Different approaches can be viable depending on the required sensitivity or runtime requirements. While some methods have been shown to efficiently recover large numbers of clusters and their members over the entire sky \citep[in particular HDBSCAN, which according to][scales better than OPTICS for large datasets]{Hunt21},
many studies focusing on specific objects use various hybrid methods to assign cluster members to known clusters. The Unsupervised Photometric Membership Assignment in Stellar Clusters \citep[UPMASK][]{KroneMartins14} is built on a k-means clustering inner loop, while its Python implementatin pyUPMASK \citep{Pera21pyupmask} supports several clustering methods. Many studies base membership assignment on Gaussian Mixture Models \citep[e.g.][]{Agarwal21,2022MNRAS.515.4685D,2023MNRAS.523.3538N}. Other successful approaches include extreme deconvolution \citep{Bovy11} used by \citet{Jaehnig21}, wavelet decompositions \citep[e.g.][]{Meingast19pisces,Furnkranz19}, support vector machines \citep[e.g.][]{Ratzenbock20,Grasser21}, self-organising maps \citep[e.g. StarGO][]{Yuan18,2022ApJ...930..103Y,Tang19,Pang20,Pang21}, significance mode analysis \citep[SigMA;][]{Ratzenbock23}, or deep neural networks \citep{2023A&A...675A..68V}.
When dealing with very small numbers of clusters, tailored modelling can be even more effective at separating cluster stars from the field populations, as shown in e.g. \citet{2022MNRAS.511.4702G} for M~37.

\subsection{Estimating cluster parameters}

In the past decades, cluster ages and interstellar extinction have traditionally been obtained via comparisons of observational colour-magnitude diagrans (CMDs) to theoretical isochrones, often performed by hand. In an era where clusters are routinely discovered by batches of several hundreds, automated methods are the key to providing homogeneous parameters for the known cluster population. 
\citet{Bossini19} applied the Bayesian code BASE-9 \citep{vonHippel06} to \Gaia photometry (supplemented with the mean parallax of the cluster) in order to obtain ages, distances, and reddening for 269 clusters. Isochrone fitting was also employed by \citet{Dias21} for 1743 objects. \citet{2022ApJ...930...44L} proposed a mixture model that is able to reproduce the presence of unresolved binaries and field star contamination, but have so far only applied their approach to a small number of clusters.

\begin{figure*}[ht!]
\begin{center} \includegraphics[width=0.99\textwidth]{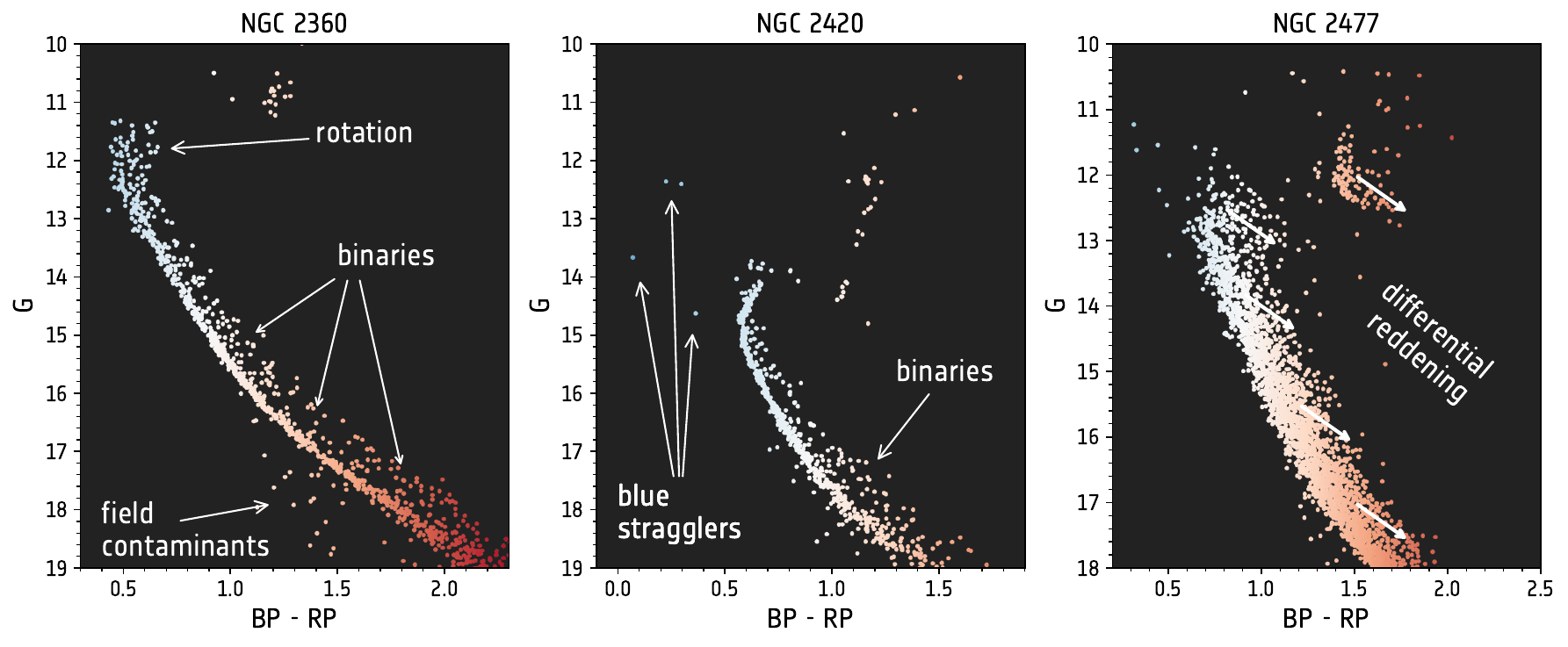} 
\caption{\label{fig:three_cmds} Colour-magnitude diagrams of three clusters with members taken from \citet{2023A&A...673A.114H}. Stellar isochrones predict that stars of the same age and chemical composition are aligned on a single line, but observed CMDs often feature deviations from this simple approximation, which must be accounted for when estimating cluster parameters. }  \end{center}
\end{figure*}

Automating isochrone fitting procedures is surprisingly not straightforward. The single sequence of a theoretical isochrone is not always a good approximation for the overall distribution of stars in an observed CMD (Fig.~\ref{fig:three_cmds}). The presence of unresolved binaries creates a second line parallel to the single-star sequence. Stellar rotation and possible age scatters can introduce a colour spread near the main-sequence turnoff. Inhomogeneous extinction across the field of view (also called differential reddening) can blur and broaden the entire CMD. In old clusters, blue straggler stars can remain brighter and bluer than the location of the turnoff. At the faint end, photometric errors affect the aspect of the CMD.
Machine learning procedures have been increasingly successful at quickly retrieving cluster parameters from their photometry, with procedures trained on synthetic data \citep[][]{Kounkel19,2023A&A...673A.114H,Cavallo23} or real clusters \citep[][]{CantatGaudin20nn}.

The median number of members in the latest open cluster catalogues is around 50, but a few hundred of the most populated clusters have 500 to several thousands of known members. \citet{Tarricq2022A&A...659A..59T}, \citet{2022AJ....164...54Z}, and \citet{2023A&A...673A.114H} all report that 90\% of the clusters have half-number radii (the radius containing half the identified members) between 2 and 6\,pc, without any obvious relation to the total number of stars. This indicates that clusters exhibit significantly different spatial densities, and dynamical states (see Sect~\ref{sec:dynamics}).

\subsection{Clusters as tracers of the Milky Way structure}

\begin{figure*}[ht!]
\begin{center} \includegraphics[width=0.99\textwidth]{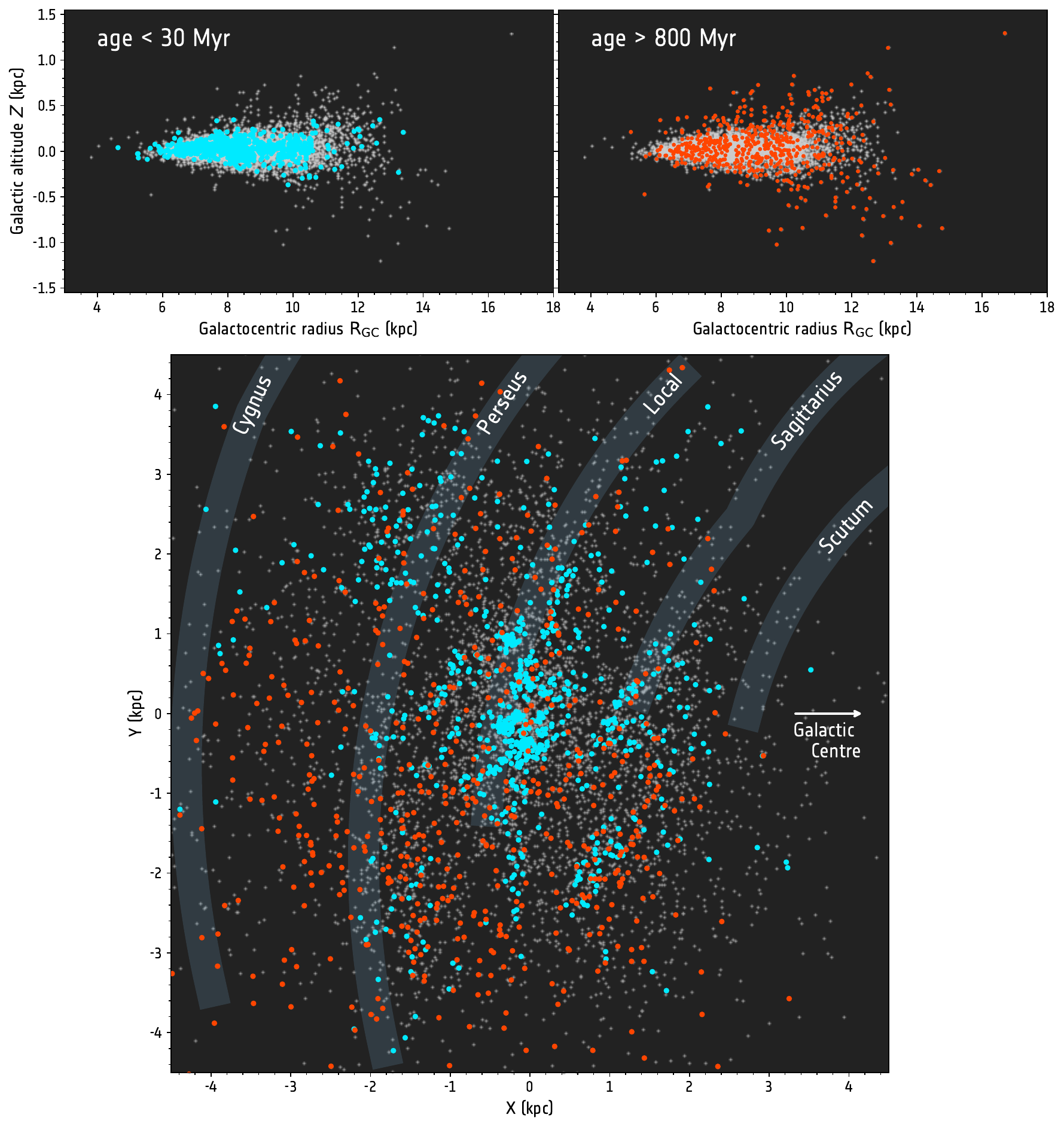} 
\caption{\label{fig:oc_census_three_panels} Spatial distribution of high-certainty clusters from \citet{2023A&A...673A.114H}, with ages from \citet{Cavallo23}. The Sun is located at R$_{\mathrm{GC}}$=8.2\,kpc and ($X$,$Y$)=(0,0). The spiral arm model is from \citet{Reid19}. }  \end{center}
\end{figure*}

Open clusters are distributed along the Galactic plane, with the majority of the youngest objects being located less than 100\,pc from the plane, while the oldest objects can be found at altitudes larger than 1\,kpc (top panels of Fig.~\ref{fig:oc_census_three_panels}). This was known before \Gaia, and is intuitively understood as the consequence of star formation taking place close to the Galactic plane (where the cold gas densities are sufficiently high) and stars subsequently dispersing as they travel through the Milky Way, gaining hotter orbits and reaching higher altitudes \citep[][]{Soubiran18,Tarricq21orbits}.

\Gaia has shown that the inner disc hosts relatively few old clusters. The median age of known clusters at Galactocentric distances R$_{\mathrm{GC}}$$<$6.5\,kpc is 120\,Myr, compared to 800\,Myr at R$_{\mathrm{GC}}$$>$12\,kpc \citep[][]{CantatGaudin20nn}. The relative lack of old clusters in the inner disc is commonly attributed to the denser environment leading to higher disruption rates. Due to large amounts of extinction, our knowledge of stellar populations in the inner disc is a lot less complete than for the Solar neighbourhood \citep[see][discussing the Galactic Extinction Horizon]{2024IAUS..377..107M}. Progress in this area is expected during this decade owing to LSST \citep{2023PASP..135g4201U} and the JAXA mission JASMINE \citep{2023arXiv230705666K}. In the long term, the proposed astrometric mission GaiaNIR \citep{GaiaNIR21} will shed light on the dense inner regions of the Milky Way.

The lack of young clusters in the outer disc is less straightforward to explain, because these regions are not devoid of young stars \citep[for instance,][have investigated the spatial distribution of Cepheids in the outer disc]{Skowron2019Sci...365..478S, Chen2019NatAs...3..320C, Lemasle2022A&A...668A..40L}.
One possible reason for the lack of known clusters could be that star formation at large galactocentric distances happens at densities that are too low to form gravitationally bound clusters. A similar idea was explored by \citet{2008Natur.455..641P} and \citet{2008Natur.451.1082K} to explain why the extension of the discs of external spiral galaxies exhibits a sharp H$\alpha$ cut off but a smoother UV profile, concluding that the outer disc does not form massive stars beyond a certain radius.
In this scenario, all the old clusters beyond a cut-off radius of R$_{\mathrm{GC}}$$\sim$12\,kpc formed in the inner Milky Way before migrating outward.

A second explanation to the lack of young outer-disc clusters could simply be the incompleteness of our catalogues. It is reasonable to believe that cluster catalogues are biased in favour of objects with high Galactic altitude, which are projected against sparser backgrounds. Cluster searches in the 5D astrometric space provided by the \textit{Gaia} can also be biased in favour of objects whose proper motion is significantly different from the surrounding field stars. Since hot orbits and large excursions from the Galactic plane are typical characteristics of old stars \citep[and old clusters, see][]{Soubiran18,Tarricq21orbits}, the cut-off radius could be explained, at least in part, by a more incomplete census of the young cluster population. There has been so far no quantitative study of the cluster selection function. 

Old clusters are survivors, while field stars are the outcome of cluster disruption. It is currently unclear to what extent the old cluster population is representative of the orbital properties of the general field population. \citet{2023A&A...679A.122V} have shown recently that the orbits of clusters older than 3\,Gyr have larger eccentricities and inclinations than field stars of the same age, suggesting that clusters are more likely to survive if their orbits allow them to spend most of their lives outside the Galactic plane. Simulations by \citet{2020MNRAS.492.4959J} also show that clusters on hot orbits have lower fractions of escapers, allowing them to survive on longer time scales.

At the far edge of the disc, distant clusters tend to be found under the Galactic plane \citep[][]{Vazquez08,CantatGaudin20nn}, following the known warp of the disc. \citet{2023arXiv230617545H} show that the variation of the line of node of the tilt with radius, as traced by clusters, is identical to the relation obtained from Cepheids. Probing the vertical structure of the outer disc with clusters is however a difficult endeavour, due to the small number of young clusters, and the large vertical scale height of the old ones.

\subsection{The spiral arms of the Milky Way}

It has long been noted that the distribution of young clusters is not uniform, and roughly follows the expected trace of the spiral arms. Until \textit{Gaia}, the uncertainties on the distances of young clusters were too large to allow for an accurate characterisation of the spiral structure in the Solar neighbourhood. The expansion of the cluster census, combined with better distance estimates owing to the parallaxes and deep photometry provide by \textit{Gaia}, have brought significant insight on the spiral arms within 2\,kpc. Rather than continuous and well-defined structures, all studies of the distribution of young clusters \citep[e.g.][]{CantatGaudin18gdr2,CantatGaudin20nn,2023A&A...673A.114H,Cavallo23} report a fragmented pattern, challenging the classical picture of a grand-design Milky Way (see Fig.~\ref{fig:oc_census_three_panels}). \citet{CantatGaudin19coin} focused on a $\sim$1\,kpc gap in the Perseus arm, showing that the lack of known clusters was not due to interstellar extinction, but to a true physical under-density of objects. The arms traced by clusters also exhibit some high-pitch structures, following those reported in the distribution of young stars such as the Cepheus spur \citep[][]{PantaleoniGonzalez21}, the spur-like structures of the second Galactic quadrant \citep{2019MNRAS.488.2158M}, or the chevrons of the inner disc \citep{Kuhn21}. Studies of the large-scale distribution of OB stars also paint a fragmented spiral \citep[][]{Poggio21,2023ApJ...947...54X}, suggesting that the Milky Way is likely to be a flocculent spiral.

The question of whether the Milky Way spiral perturbations are global and stationary \citep{Lin64} 
or local and transient \citep{Toomre64} has been a matter of debate for decades \citep[see reviews by][]{Dobbs14,Shu16,Xu18,Shen20}.
Using young clusters ($<$30\,Myr) with precise 3D positions and velocities, \citet{CastroGinard21spiral} have shown that the four classical spiral arms have independent pattern speeds, all of them close to the corotation of the disc. These findings support the idea that they are short-lived structures, rather than perennial Galaxy-scale density waves. Interestingly, their result is at odds with \citet{Dias2019MNRAS.486.5726D} who included slightly older clusters (up to 50\,Myr) and report no significant difference in pattern speed between the arms. 

Recent studies mapping the chemical distribution of field stars by \citet{Poggio2022A&A...666L...4P} and \citet{Hawkins2023MNRAS.525.3318H} have observed that giants associated with the spiral arms exhibit higher metallicities than those located in the interarm regions. It is currently not possible to confirm this result using young clusters, due to the lack of available metallicities for these objects.

\section{Young clusters, associations, moving groups} \label{sec:young}

The \textit{Gaia} parallaxes have enabled detailed studies of the 3D distribution of young stars in the Solar neighbourhood, revealing a wide range of sizes and densities within 150\,pc \citep[][using DR1 data]{Gagne18dr1,Gagne2019clrp.2020....1G}, several hundreds of parsecs \citep[][with DR2]{Zari18,Kounkel19,Kerr21}, and beyond a kiloparsec with DR3 data \citep{Zari2021A&A...650A.112Z,McBride21,2022A&A...664A.175P,2022ApJ...931..156P,2023arXiv230608150K}.

In this section we discuss some key ideas that \textit{Gaia} has introduced or solidified in the recent years concerning the structure and origin of young stellar aggregates.

\subsection{Loose associations are (usually) not expanding clusters}

During the early phases of star formation, radiation pressure from hot stars and supernovae can rapidly strip a proto-cluster from its gas, causing the cluster to become supervirial and its stars to quickly disperse. While this scenario can happen in the Milky Way \citep[e.g. $\lambda$~Orionis,][]{Kounkel18},
it was reported even before \textit{Gaia} that some associations are not rapidly expanding \citep{Wright2016MNRAS.460.2593W,2017MmSAI..88..850W,2017sfcc.confE..63W} a result allowed by theoretical models \citep[e.g.][]{Kruijssen12,Kruijssen12dynamical}. The \textit{Gaia} data indicates that not just some, but the majority of loose stellar aggregates in the Solar neighbourhood are in fact expanding slowly \citep{Kuhn18,Ward18,Melnik20,Ward20}.

As associations are barely supervirial, they retain some of their initial structure and provide clues on the conditions of star formation. Perhaps the best example is provided by the Scorpius-Centaurus association (Sco-Cen), the most nearby and the most studied major association. While Hipparcos studies of Sco-Cen were able to identify a few hundred members \citep{deZeeuw99}, the \textit{Gaia} data has allowed the detection of low-mass stars bringing the census to 15,000 members \citep{Damiani18,Gagne18dr1,Wright18scocen,Damiani19scocen,Roser18,Goldman18,Luhman18uppersco,Luhman20,Squicciarini21,Luhman22,Zerjal23,Briceno2023MNRAS.522.1288B,Ratzenbok23scocen,Ratzenbock23}. The recent study by \citet{Ratzenbock23} reports the identification of 37 distinct and coeval clusters within the association, reaching densities as low as 0.01 sources\,pc$^{-3}$, and with relative velocities of 0.5\,km\,s$^{-1}$. The authors also show a continuous age gradient through the entire structure, hinting at sustained star formation across $\sim$15\,Myr.
\citet{Briceno2023MNRAS.522.1288B} investigated the space-kinematics-age structure of Sco-Cen, and propose that four massive stars have shaped the present structure of the association through stellar winds and supernovae. 

Despite being gravitationally unbound, stellar groupings can remain spatially coherent over hundreds of Myr \citep[e.g.][]{Meingast19pisces,Kounkel19, Gagne2021ApJ...915L..29G,Gagne2023ApJ...945..119G}. Systems that are neither dense nor young are often referred to as moving groups.
As a complete overview of studies of stellar associations performed with \textit{Gaia} data would be impractical, we refer the reader to  Section 3 of \citet{2022Univ....8..111C} for an almost exhaustive review of association studies up to December 2021, and to \citet{Krumholz19} (Sect. 3.4.2 and 3.5.2) and \citet{Wright20} (Sect. 5.3) for a discussion of the origin of these structures.

\subsection{Single events of star formation can span hundreds of parsecs}

The ability to identify low-mass members of associations and young clusters has revealed that some coeval and co-moving superstructures can be traced over hundreds of parsecs. Investigations of the Vela~OB2 association with \textit{Gaia} have increased its number of known members from $\sim$200 \citep{deZeeuw99} to over 14,000 \citep{Franciosini18,Armstrong18,Beccari18,CantatGaudin19vela,CantatGaudin19vela2,Beccari20,Pang21} and revealed a high degree of spatial and kinematic substructure. \citet{CantatGaudin19vela2} have shown that the clusters NGC~2547, NGC~2451B, Collinder~140, Collinder~135, and UBC~7 are part of a continuous, co-moving alignment of coeval ($\sim$35\,Myr) stars spreading over 200\,pc. \citet{Beccari20} discovered another cluster associated with this structure, showing that this kinematically cold family currently spans at least 260\,pc. The young age and the small velocity dispersion within the group indicate that this morphology cannot be due to tidal disruption, and therefore reflects the filamentary structure of the parent molecular cloud, which rapidly collapsed into a coeval population.

\subsection{Young massive clusters: a distinct mode of star formation?}

The inner regions of the Milky Way host a handful of very massive clusters ($>10^4$\,M$_{\odot}$) that contain a dense population of massive stars, and are strongly affected by extinction and reddening. 
Westerlund~1 is a $5 \times 10^4$\,M$_{\odot}$ cluster \citep[see][and references therein]{Andersen2017A&A...602A..22A}.
\textit{Gaia}~DR2 studies of this cluster placed its distance between 2.5 \citep{2020MNRAS.492.2497A} and 4\,kpc \citep{2019MNRAS.486L..10D}. Recent studies support a distance of 4\,kpc, albeit still with a significant uncertainty \citep{2022MNRAS.517.3749R,2022MNRAS.516.1289N,2022A&A...664A.146N}.
\citet{2021ApJ...912...16B} estimate that the pre-main sequence stars in the cluster are 7\,Myr old, while the W13 eclipsing binary is 5\,Myr at most. \citet{2022MNRAS.517.3749R} propose an age of 7\,Myr for the eclipsing binary W36B and OB stars, but an older age of 10,7\,Myr for the red supergiants, a result confirmed by \citet{2022MNRAS.516.1289N}. \citet{Beasor2023ApJ...952..113B} also present evidence in favour of a multi-age cluster, resulting from several bursts of star formation. \citet{2022A&A...660A..89R} report that the binary fraction in Westerlund~1 is at least 40\%. 

Although the cores of young massive clusters cannot be easily observed directly at optical wavelengths, the \textit{Gaia} data has been used to study their surroundings, and in particular the population of runaway massive stars expelled from these dense regions. Evidence for recent and non-isotropic ejection in Westerlund~2 \citep{Drew2018MNRAS.480.2109D,Zeidler21} supports the scenario of massive clusters growing via mergers. However, \citet{Drew19} remark that in NGC~3603 the spatial distribution of escapers could be better explained by a cluster core collapse rather than the merging of fully-formed clusters. We refer the interested reader to \citet{2014prpl.conf..291L} for a (pre-\textit{Gaia}) overview of observational and theoretical studies of young massive clusters in the Milky Way and the Magellanic Clouds, and to the study of Westerlund~1 and 2 by \citet{Guarcello2023arXiv231208947G} for a summary of the open questions regarding these objects. In Sect.~\ref{sec:dynamics} of this review, we further discuss studies of stars escaping for clusters.

While they are not expected in the Solar neighbourhood, supermassive clusters are expected to be formed in the inner regions of spiral galaxies \citep{2023MNRAS.524..555A}, from hierarchical merging of smaller clusters \citep[e.g. simulations by][]{Howard2018NatAs...2..725H,Rieder22,Guszejnov2022MNRAS.515..167G,2022MNRAS.517..675D,2023MNRAS.524..555A,2023MNRAS.521.1338C}. The structure and kinematics of these obscured regions are difficult to probe with \textit{Gaia} data, but represent an obvious science case for the near-infrared space-based astrometric mission GaiaNIR.

\section{Estimating cluster ages} \label{sec:ages}

The ages of most clusters, and possible departures from a coeval star formation scenario, can be estimated by comparing the distribution of their stars in a colour-magnitude diagram with theoretical isochrones. The accuracy and precision of the resulting age depends on prior knowledge of the cluster metallicity and extinction, but also on the number of cluster members in key evolutionary phases. For clusters older than $\sim$100\,Myr, the colour of the bluest main-sequence stars combined with the presence of red giants provide a simple constraint on the cluster age. For clusters younger than $\sim$40\,Myr, the presence of pre-main-sequence stars can be an age indicator, provided that the cluster is not strongly obscured by extinction, and nearby enough for its low-mass population to be observable. In this section we review various  indicators other than photometric isochrone fitting used by recent studies to estimate the ages of stellar clusters and associations.

\subsection{The Lithium depletion boundary}

Stars burn Li in their cores, but not in their outermost layers. For convective stars, the mixing of material progressively depletes Li throughout the entire star, while for stars of F and earlier types the surface Li depletion remains minimal. The Li depletion boundary (LDB) technique relies on establishing the age-dependent mass (or luminosity) at which low-mass stars have not yet depleted their Li. In practice, it requires to measure the equivalent width of the Li~I 6708\,$\AA$ absorption line in the stellar spectra of cluster stars, at resolution greater than 3000 \citep{Duncan1983ApJ...271..663D,Stauffer1998ApJ...499L.199S,Jeffries2009MNRAS.400..317J,Jeffries2013MNRAS.434.2438J}

Ages obtained fron the LDB method are sometimes discrepant from those obtained from photometric isochrone fitting \citep[e.g.][]{Jeffries2017MNRAS.464.1456J,2022A&A...664A..70G}. Based on Li abundances collected by the \textit{Gaia}-ESO Survey \citep{Randich13},  \citet{2022A&A...659A..85F} propose a set of pre-main-sequence models that include radius inflation due to the presence of star spots and magnetic activity. They also point out that for the clusters younger than $\sim$20\,Myr, their models require more spot coverage on the low-mass stars to fit the observations.

Lithium abundances have also been shown to be related to stellar rotation, with fast rotators preserving their Li reserves over longer time scales \citep[see][and references therein]{Jeffries2021MNRAS.500.1158J}. \citet{2022MNRAS.513.5727B} have studied five clusters with ages ranging from 5 to 125\,Myr. They show that the apparent LDB age spread of NGC~2264 is in fact a rotation rate spread. On the other hand, they found that the $\gamma$~Velorum cluster exhibits no correlation between Li and rotation, hinting at a true age spread for this cluster. \citet{2023ApJ...952...71S} have shown that even in the 420\,Myr old cluster M~48, fast-rotating G-K stars have larger Li abundances than their slowly rotating counterparts. \citet{2023A&A...674A.157T} have shown that this relation also exists for Li-rich giants. \citet{2023MNRAS.523..802J} also provide empirical models\footnote{Empirical AGes from Lithium Equivalent widthS (EAGLES); \url{https://github.com/robdjeff/eagles} }, calibrated on 6200 stars in 52 open clusters with ages ranging from 2\,Myr to 6\,Gyr.

\subsection{Gyrochronology}

The study of how stellar rotation spins down with time has led to the emergence of gyrochronology, which uses rotation periods as a proxy for age \citep{Barnes03,Douglas14,Douglas16,Meibom15}. A great advantage of colour--period diagrams over isochrone fitting is the possibility to obtain ages for main-sequence stars. This field has benefited from a tremendous boost enabled by \textit{Gaia} astrometry (allowing for pure membership lists) and the data collected by the space-based missions CoRoT \citep{Auvergne09}, Kepler \citep{Borucki10}, K2 \citep{Howell14}, and the Transiting Exoplanet Survey Satellite (TESS, \citep{Ricker15}), but gyrochronology can also be performed with ground-based observations \citep{GodoyRivera21}.

A large number of recent studies have been focusing on well-known clusters in order to better understand the relation between rotation, age, and stellar type. Gyrochronological investigations have been performed for the Hyades \citep{Douglas19}, Pleiades \citep{Douglas19}, NGC~2516 \citep{Fritzewski20,Bouma21}, Ruprecht~147 \citep{Gruner20,Curtis20}, NGC~6811 \citep{Curtis19ngc6811}, NGC~3532 \citep{2021A&A...652A..60F, 2021A&A...656A.103F}, M~67 \citep{2023A&A...672A.159G}, or NGC~6709 \citep{2023A&A...673A.119C}.

Gyrochronology has also been successfully employed to constrain the ages of clusters that are either too young or too sparse to host evolved stars to use as age markers. \citet{Curtis19pisces} have shown that the age of the Pisces-Eridanus stream \citep{Meingast19pisces} is comparable to that of the Pleiades. \citet{2022A&A...657L...3M} and \citet{2022AJ....164..115N} show that the moving group X is coeval with NGC~3532 ($\sim$300\,Myr) and therefore not associated with the much older Coma Berenices. \citet{2023A&A...674A.146P} have studied rotation periods NGC~2477 (a cluster strongly affected by differential reddening) and reduced the uncertainty on its age from 0.3 to 0.1\,dex. \citet{2023MNRAS.522.4894F} have shown that the sparse cluster ASCC~123 is coeval with the Pleiades. NGC~2281 is another cluster for which uncertain age estimates range from $\sim$200 to 700\,Myr. \citet{2023A&A...674A.152F} estimate a gyrochronological age of 435$\pm$50\,Myr. Recently, \citet{Fritzewski2023arXiv231018426F} were able to constrain the age of the sparse UBC~1 from rotation periods and the identification of a single star exhibiting gravity-mode pulsations.

\citet{2023AJ....166...14B} have exploited rotation periods to select a sample of stars coeval with the $\sim$80\,Mr $\alpha$~Per association, and recover the full extent of the complex defined as Theia~133 by \citet{Kounkel19}. The possibility to identify young stars from their rotation periods \citep[as young as 40\,Myr;][]{Douglas2023arXiv231117184D} opens exciting possibilities for the study of the spatial and kinematic structures of young stellar complexes.

While many studies rely on visual comparisons between period distributions, tools have been developed to simplify the use of gyrochronological data.
\citet{2019AJ....158..173A} propose a Python package to simultaneously derive ages from photometry and rotation periods\footnote{\url{https://github.com/RuthAngus/stardate}}, and
\citet{2023ApJ...947L...3B} make tools available for the computation of empirical rotation--temperature relations\footnote{\url{https://github.com/lgbouma/gyro-interp}}. Finally, \citet{2023arXiv230708753V} have recently explored the use of normalising flows to build a data-driven probabilistic model extracting stellar ages.

\subsection{Ages from internal kinematics}

The possibility to resolve the 3D spatial and 3D kinematic distribution of young clusters and associations has allowed comparisons between their isochronal ages and estimates from dynamic traceback of their member stars. This approach can be especially useful for sparse objects offering little age constraints in their CMDs. For instance, \citet{MiretRoige20betapictoris} obtained a traceback age of 18.5$\pm$2\,Myr for $\beta$~Pictoris, a cluster for which literature estimates range from 10 to 40\,Myr. The estimates are however subject to systematics. \citet{2023ApJ...946....6C} show that a possible, radial velocity shift of 0.6\,km\,s$^{-1}$ affects the traceback age of $\beta$~Pictoris by $\sim$2\,Myr, and the presence of stars with uncertain membership status by 3\,Myr. \citet{2023MNRAS.520.6245G} found the dynamical age of Tucana-Horologium ($\sim$40\,Myr) to be consistent with both isochronal and Li ages.

\citet{2022A&A...667A.163M} remark that the dynamical age of Upper Scorpius appear younger than the age estimated from its CMD. Further investigation of this idea by \citet{2023NatAs.tmp..245M} for six young stellar associations shows a systematic difference of $\sim$5\,Myr, which the authors suggest is the typical time a star remains bound to its siblings before moving away: the isochrone age of a group indicates the time its stars were born, while the dynamical traceback age indicates when it started expanding.  \citet{2023arXiv231108363P} proposed and validated a method based on the expulsion age of individual stars, showing that the oldest age (corresponding to the first star to leave the cluster) generally provides a better match to the isochronal age than the traceback method.

While the above methods are only applicable to young unbound associations, \citet{2022ApJ...925..214D} propose an approach based on the observed tilt angle between the orbital direction and the tidal tails of dissolving clusters, which can be predicted analytically. The authors estimate that such age estimates can have a precision of 10 to 20\% for clusters younger than $\sim$300\,Myr.

\section{Internal kinematics and dynamics} \label{sec:dynamics}

A gravitationally bound cluster finds itself in a situation of temporary but unstable equilibrium. While its stars orbit each other following their collective gravitational potential, they also feel the presence of the gravitational potential of the entire Milky Way. Stars that reach the cluster's escape velocity (due to stochastic 2- or 3-body encounters) may become lost to the Galactic field population. Stars preferentially escape through the Lagrange points of the cluster \citep{2000MNRAS.318..753F,Gieles08tidally,PortegiesZwart10}, creating a leading and a trailing tidal tail. While the tidal tails of globular clusters have been studied for decades \citep[e.g.][]{Odenkirchen2001ApJ...548L.165O,Belokurov2006ApJ...637L..29B}, the study of open cluster tidal tails only became possible in the \textit{Gaia} era due to the relatively low density of these objects, and the much higher background and foreground contamination.

The elongated morphology of the Hyades was first reported by \citet{Reino18} using \textit{Gaia}~DR1 data. Its tidal tails were later mapped in greater detail with DR2 data \citep{Lodieu19,Roeser19hyades,Meingast19,Oh20}. With DR3 astrometry, \citet{Jerabkova21} traced the Hyades tidal tails over a distance of 800\,pc. Other nearby clusters exhibit prominent tidal tails: NGC~2632 \citep{Roeser19praesepe}; Ruprecht~147 \citep{Yeh19}; M~67 \citep{Carrera19m67}; Coma Berenices \citep{Tang19}; NGC~752 \citep{Bhattacharya21ngc752,2022MNRAS.514.3579B}; UBC~274 \citep{2022A&A...664A..31C}; COIN-Gaia~13 \citep{Bai2022RAA....22e5022B}. \citet{2021ApJ...912....5H}, \citet{Hu21decoding}  and \citet{Tarricq2022A&A...659A..59T} report that nearly a hundred nearby clusters have an elongated morphology, with a preferential elongation parallax to the Galactic plane, consistent with the expectation of tidal disruption. 
\citet{2022MNRAS.517.3525B} found tidal tails in 20 clusters, and a total of 46 clusters with stars outside their tidal radius.

\begin{figure*}[ht!]
\begin{center} \includegraphics[width=0.99\textwidth]{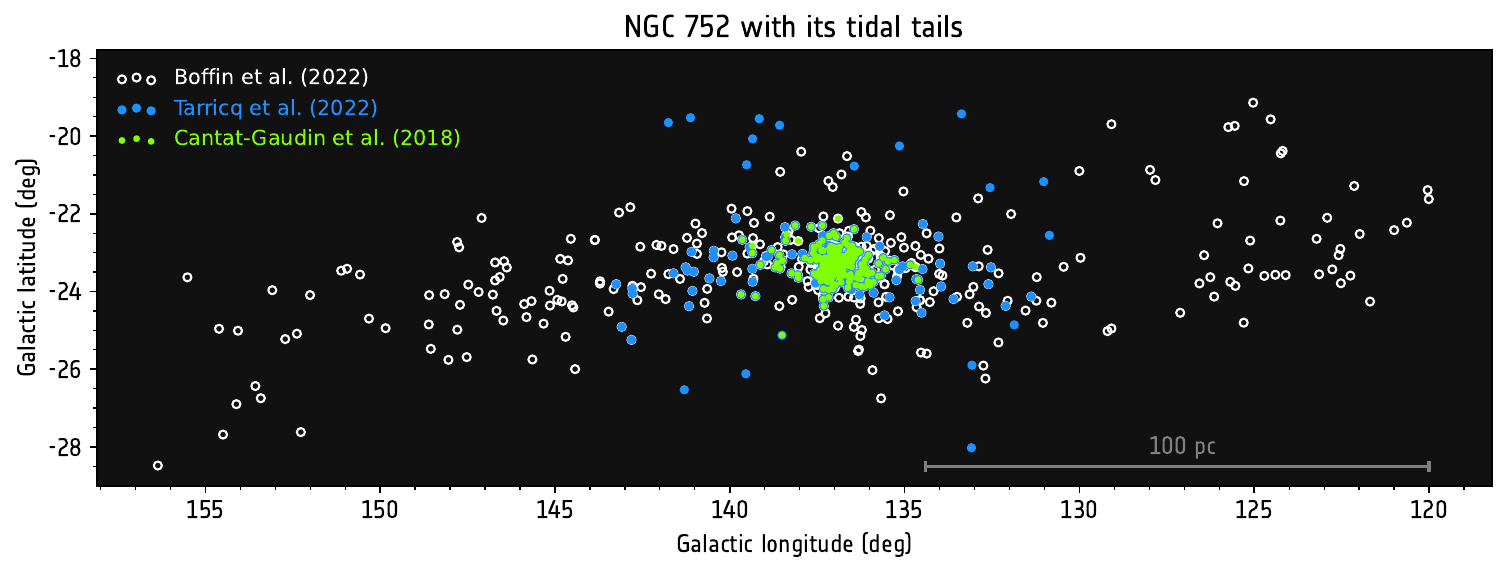} 
\caption{\label{fig:ngc_752} Members of NGC~752 identified by \citet{CantatGaudin18gdr2} by clustering in astrometric space (green), by \citet{Tarricq2022A&A...659A..59T} modelling an elongated structure (blue), and by \citet{2022MNRAS.514.3579B} with a convergent point method. }  \end{center}
\end{figure*}

In Fig.~\ref{fig:ngc_752} we show how \textit{Gaia} has increased the known extent of the tidal tails of NGC~752, which according to \citet{2022MNRAS.514.3579B} span up to 260\,pc (over 25 degrees on the sky). Studying objects with large angular sizes requires to properly account for projection effects, both in their spatial and kinematic distribution. This can be done with convergent-point methods \citep[e.g.][]{Meingast21,2022MNRAS.514.3579B,2023A&A...678A..75Z}, or by converting the observed \textit{Gaia} astrometry into Cartesian coordinates \citep{Gagne18banyan,2022ApJ...939...94M}, or even performing clustering analyses on transformed quantities such as action angles \citep[e.g.][]{Coronado20,2022ApJ...928...70C,2024ApJ...961..113F}.

\citet{2022MNRAS.517.3613K} have shown that the tidal tails of the Hyades are significantly asymmetrical, with a leading tail more populated than the trailing tail. \citet{2023A&A...671A..88P} suggest this is also the case in NGC~2632, Coma Berenices, and NGC~752, although the results are statistically less robust for these more distant objects. Interestingly, \citet{Thomas2018A&A...609A..44T} have theorised that asymmetric tidal tails naturally arise from MOdified Newtonian Dynamics \citep[MOND;][]{Milgrom1983ApJ...270..371M,Sanders2002ARA&A..40..263S,2005idm..conf...49M,Famaey2012LRR....15...10F}. While \citet{2023A&A...671A..88P} do consider that the observed asymmetry of the Hyades could be due to local bumps in the Milky Way potential or to external perturbations, they point out the need for investigations of the morphology of tidal tails in a larger number of clusters. \citet{2022MNRAS.517.3613K} point out that MOND also predicts clusters older than 200\,Myr should experience a spin-up opposite to their orbital angular momentum, a prediction that the current data is not able to verify yet. The low-acceleration regime surrounding dissolving clusters in the Galactic disc provides a good opportunity to test this theory.

Detecting the signal of open cluster rotation is challenging, even in the \textit{Gaia} era. Evidence for clear rotation was found in NGC~2632 by \citet{Loktin2020AN....341..638L} and \citet{2022ApJ...938..100H}, with a significant tilt of 41$\pm$12$^{\circ}$ with respect to the Galactic plane. \citet{2023A&A...673A.128G} report a rotational signature in the plane of the sky for eight more clusters, detected from \textit{Gaia} proper motions. By supplementing \textit{Gaia} proper motions with ground-based spectroscopy, \citet{Kamann19} reported evidence for rotation in NGC~6791, but not in NGC~6819 (two clusters older than 1\,Gyr). The origin of rotational patterns in clusters can be multiple: it can be inherited at birth, imprinted by interactions with massive structures, or due to the long-term action of tidal forces. These different mechanisms should result in distinct patterns (for instance, strong interactions would preferentially induce solid-body rotation). Given the small number of clusters with known rotational signatures, we are currently far from being able to construct the empirical relations between cluster age and rotation.

Interpreting the observational clues about the dynamical state of clusters is made even more difficult by the mechanism of mass segregation. The equipartition of kinetic energy causes more massive stars to orbit closer to the centre of the cluster, with a lower velocity than low-mass stars \citep{1969ApJ...158L.139S}. This phenomenon has been observed in many Galactic star clusters, and causes the low-mass members to be lost to the Galactic field at higher rates than the high-mass members. In a study of the strongly mass-segregated Hyades, \citet{2022MNRAS.512.3846E} show that Milky Way open clusters can in fact never be near energy equipartition, due to the combined influence of the Galactic tidal field and the evaporation driven by mass segregation.

Mass segregation is common in open clusters, and Ruprecht~147 \citep{Yeh19}, Czernik~3 \citep{Sharma20}, and ASCC~92 \citep{2023MNRAS.519.6239P} have even been shown to be heavily depleted in low-mass members. Since cluster membership can only be established up to a certain magnitude and within a given radius, estimates of the total mass of a cluster are often based on extrapolations of the known stellar content, and therefore affected by significant uncertainties. 
\citet{2022MNRAS.516.5637E} determined the mass function and the dynamical parameters of 15 nearby clusters, and show a correlation between the slope $\alpha$ of the mass function and the quantity $\log(t_{age} / t_{rh})$, which is the ratio of the age to relaxation time. The resulting sequence is analogous to the corresponding relation obtained for globular clusters, but with a steeper mass function. The authors propose that this difference could be primordial, and that Galactic clusters are born with a steeper initial mass function than globular clusters.
\citet{2023MNRAS.525.2315A} provide mass estimates for 773 clusters (from 100 to 2000\,M$_{\odot}$), and find no apparent dependence between mass segregation and age. However, \citet{2023MNRAS.522..956A} report that the core radius of clusters appear to decrease with age, and that the older clusters are those that tend to underfill their Jacobi radius. They also report that inner-disc clusters appear dynamically older and to have experienced more mass loss than in the outer disc. Comparing the dynamical properties of clusters in different Galactic environment is however difficult, especially given the uncertainty in establishing the membership of inner-disc clusters affected by extinction and projected against a dense background field population.

Understanding the dynamical evolution of star clusters requires some insight on their binary population. \citet{2023A&A...675A..89D} studied the binary fraction in 202 clusters, and found that high-mass stars are more likely to be in binary systems. They report a typical binary fraction of 18\%, but with significant variations, ranging from $\sim$10 to 80\% throughout the sample of clusters studied. \citet{2023A&A...672A..29C}, on the other hand, found a flat relation between binary fraction and primary mass in a study of 78 clusters. The authors also report that in clusters older than several Gyr, the binaries appear more centrally concentrated than the single stars. \citet{2023arXiv230816282C} present a method to account for the presence of photometric binaries in colour-magnitude diagrams, and remark that the binary fraction increases with age in their sample of six clusters. \citet{2023A&A...675A..89D} discuss trends of binary fraction with age appearing in their studies, and warn that they are most likely caused by selection effects in the cluster membership. They also report no apparent correlation between binary fraction and location in the Milky Way. In the neaar future, astrometric time series provided by \textit{Gaia} might help provide further constraints on the binary content of star clusters.

Another relevant piece of information in describing the dynamical state of a cluster is whether it hosts stellar-mass black holes. Comparing \textit{Gaia} observations to N-body simulations, \citet{2023MNRAS.524.1965T} show that the half-mass radius of the Hyades is 50\% too large for its total stellar content, and more consistent with a system retaining two or three black holes.

\section{Stellar evolution and chemistry} \label{sec:spectro}

\subsection{Clusters and the stellar bestiary}

The clean CMDs obtained from pure membership lists and millimag \textit{Gaia} photometry allow for robust calibrations of stellar evolution models \citep[e.g.][]{2022ApJ...938..125M,2023MNRAS.518..662B,2023AJ....165..108B,2023arXiv230603132B}, and for the identification of stars in specific phases of their evolution.

Some clusters exhibit an extended turnoff. While it was sometimes proposed that age spreads in open clusters could cause a spread in colour near the turnoff point, recent studies have shown that the redder side of extended turnoffs often corresponds to fast rotators, whereas bluer stars are slower rotators \citep{Marino18,Bastian18,Cordoni18,Lim19,2022ApJ...938...42H,2023MNRAS.524..108G}. 
Rotation is essential but not unique in explaining broadened turnoffs. In the Magellanic Clouds, this phenomenon has also been linked to significant age spreads \citep[e.g.][]{Goudfrooij14} and stellar variability \citep{Salinas16}.
We note that in Stock~2, \citet{AlonsoSantiago21} observe no significant different in rotation rates, and attribute the extended turnoff to differential reddening.

Blue straggler stars (BSSs) are bright stars whose colour is bluer than the main sequence turnoff of old clusters. The number of known BSSs is now over 1500, in over 300 different clusters \citep{2023A&A...672A..81L}. The main idea invoked to explain these objects is that they have recently gained mass, either via merger or mass transfer from a binary companion. \citet{Vaidya20} and \citet{Rain21} observed that the spatial distribution of BSSs is not more concentrated than that of other high-mass stars. \citet{Leiner21} investigated sixteen old open clusters and found that standard population synthesis produces too few BSSs with respect to observations, proposing updates to mass-transfer prescriptions. Ultraviolet studies by \citet{2022MNRAS.516.5318P}, \citet{2022MNRAS.511.2274V}, \citet{2022MNRAS.511.2274V}, and \citet{2023arXiv230615396J}, were able to confirm a large number of BSSs as post-mass-transfer binary objects.
Interestingly, \citet{2023MNRAS.518L...7R} observe a double BSS sequence in Berkeley~17, which they warn might be coincidental.

Star clusters are a privileged ground for the observation of white dwarfs (WDs). Since robust estimates of the cluster age are often available, the mass of the WD progenitors can be constrained much more accurately than for field WDs \citep{Si18,Canton21,Heyl2022ApJ...926..132H}. Multiple studies have searched for WDs associated with clusters \citep{GentileFusillo19,Prisegen21,Richer21}. \citet{2023arXiv230701337P} remark that very few WDs with progenitors more massive than 5\,M$_{\odot}$ are known to reside in clusters, suggesting that more massive objects are subjected to strong kicks and ejected from clusters.

\subsection{Chemical abundances of stellar clusters}

In the past ten years, spectroscopic surveys have been collecting data enabling detailed chemical analyses of stellar populations in the Milky Way. Some of the largest observational campaigns, such as the Gaia-ESO survey \citep{Randich13}, APOGEE \citep[Apache Point Observatory Galactic Evolution Experiment;][]{Majewski17}, RAVE \citep[Radial Velocity Experiment;][]{Steinmetz2020}, LEGUE \citep[LAMOST Experiment for Galactic Understanding and Exploration;][]{Deng12}, or GALAH  \citep[GALactic Archaeology with HERMES;][]{DeSilva15} have selected cluster stars among their targets. Smaller programmes dedicated to clusters are also currently ongoing, such as OCCASO (Open Cluster Chemical Abundances from Spanish Observatories) survey \citep{Casamiquela16,Carrera2022,CarbajoHijarrubia24}, SPA \citep[Stellar Population Astrophysics;][]{Zhang21}, OSTTA\citep[One Star to Tag Them All;][]{Carrera2022b}, or BOCCE \citep[Bologna Open Clusters Chemical Evolution;][]{Bragaglia07}. 

In the pre-\textit{Gaia} era, the selection of probable members for spectroscopic follow-up was sometimes a tedious task \citep[e.g.][]{2022A&A...659A.200B} and could lead to the selection of a large fraction of non-members \citep{Kos18}. The \textit{Gaia} astrometry offers a more secure membership estimate down to the faint magnitude reached by the current generation of multi-object spectrographs such as WEAVE \citep{JinWEAVE} and 4MOST \citep{Dalton16,2023Msngr.190...13L}.

The \textit{Gaia} spacecraft itself has limited spectroscopic capabilities, through its Radial Velocity Spectrograph  \citep[RVS;][]{Cropper18,Katz19} observing in the 845--872\,nm wavelength range where Fe and prominent Ca~II lines are visible in stellar spectra. While this instrument was mainly designed to allow for radial velocity measurements, \textit{Gaia}~DR3 came with the publication of stellar parameters and chemical abundances for 5.6 million stars \citep{RecioBlanco2023}.
\citet{2023A&A...674A..38G} explored the chemical properties of stars belonging to open clusters in the context of the thin disk from Gaia DR3 chemistry only.
Their sample contained 503 clusters older than 100\,Myr, and was limited to Galactocentric radius of 12\,kpc due to the limiting magnitude of RVS.
This sample, though with larger overall uncertainties than high-resolution studies, is several times larger than any previous sample and allowed to investigate the radial [M/H] and [$\alpha$/H] gradient in the Galaxy, and their evolution in age bins.
Mean radial [M/H] gradients seem compatible with previous literature studies from open clusters with much less statistics \citep[e.g.][]{Spina21,Casamiquela2019}.
Interestingly, the evolution of the gradient as a function of age investigated by \citet{2023A&A...674A..38G} shows a steepening with age, the opposite of the results reported by \citet{Anders17} and \citet{Spina21}.

High-resolution ($R>20,000$) spectroscopic observations are essential to retrieve abundances with precision better than $0.05$\,dex, which is usually desired to perform Galactic archaeology studies \citep[e.g.][]{Jofre2019}.
From APOGEE data, \citet{Donor20} studied the radial abundance gradients of 16 chemical elements, using a sample of 128 open clusters spanning a large range of Galactocentric radii ($6<R_{GC}<18$ kpc).
A change of slope in the radial metallicity gradient in the outer disk is confirmed though the exact radius at which this happens depends on the distance catalog used.  On the other hand, \citep{Magrini2023} used Gaia-ESO survey data of 62 clusters between $6<R_{GC}<20$ kpc to investigate the evolution of the chemical gradients for 24 chemical elements, they find to be very limited in time, indicating a slow and stationary evolution of the thin disk. For a recent overview of studies of abundance gradients in the Milky Way using star clusters as tracers, we refer the reader to \citet{2022Univ....8...87S}.

In clusters, the precision of high signal-to-noise spectra ($>100$) can be improved to 0.01--0.02\,dex by using a cluster member as a reference to derive differential abundances for the other similar stars at the same location in the HR diagram \citep[e.g.][]{Casamiquela2020}.
At this precision it becomes possible to study internal chemical homogeneity of clusters \citep{Liu2016,Liu2016M67}, which can be related to the internal mixing level of the proto-cluster cloud, or the effects of planetary formation.
\citep{Manea2022} studied the chemical abundances derived from GALAH of five of the large stellar strings identified by Gaia \citep{Kounkel19}, which seem to be in general chemically homogeneous.
However, chemical studies with larger statistics are needed to clarify which of these filamentary structures are real physically related objects \citep[see also][]{Zucker2022}.

Clusters are reference objects due to their large number of stars presumably sharing their age and chemical composition.
Deriving abundances of cluster members from automatic pipelines, is a typical way of assessing the internal precision (from the dispersion among its members) and the systematic uncertainties using stars of different stellar parameters \citep[e.g. dwarfs vs giants;]{Jofre2019}.
Star clusters are also good laboratories to test methods and assumptions.
A good example is the chemical tagging idea \citep{Freeman02}, which states that it should be possible to use chemical information to tag stars that formed within the same proto-cluster cloud, even in the cases where the cluster no longer exists. Since some clusters are known to exhibit unique abundance patterns \citep[e.g. UBC~274 with an overabundance of neutron-capture elements][]{2022A&A...664A..31C}, chemical tagging with high-precision abundances should be feasible in theory.
Studies in the recent years have shown that this idea seems to have low chances of success in practice, in the first place due to internal chemical differences found in member stars, as we have discussed in the previous paragraphs.
Additionally, it is now more and more evident that the chemical patterns of known clusters in the thin disk have a large overlap \citep{BlancoCuaresma2015,Casamiquela21tagging}, thus, making most of the clustering techniques to fail to recover members of known clusters from blind chemical searches \citep[see also discussion in][]{Spina21}.
The only study in the literature to attempt to identify new clusters in large catalogs using chemistry only is \citet{PriceJones2020}, from APOGEE data.
However, the inability of their method to recover any of the known clusters in their sample, and the fact that real cluster members show a larger abundance dispersion than the identified new groups, highlights the difficulty of the task.

\section{Summary and conclusions}

\textit{Gaia}'s microarcsecond astrometry has already revolutionised Milky Way astronomy, and perhaps the studies of Galactic clusters more than any other field. The ability to discover and confirm new clusters and to estimate their main parameters allows to map the stellar content our Galaxy with unprecedented accuracy within 4\,kpc of the Sun. 
The possibility to establish reliable lists of cluster members down to faint magnitudes is invaluable for follow-up studies, and in particular the construction of target lists for large ground-based spectroscopic surveys.

We are now entering an era where star clusters can no longer be treated as individual data points, but as fluid objects in interaction with the Milky Way's gravitational potential. However, only a small fraction of the known clusters can be resolved kinematically, and probing their dynamical state is not straightforward. Large regions of the Milky Way remain obscured by dust and difficult to investigate by \textit{Gaia}. While we can expect improved data from the upcoming \textit{Gaia} data releases, it is probably safe to assume that no single instrument will have, in the near future, a transformative impact comparable to \textit{Gaia}.

\section*{Acknowledgements}
This work has made use of data from the European Space Agency (ESA) mission {\it Gaia} (\url{https://www.cosmos.esa.int/gaia}), processed by the {\it Gaia} Data Processing and Analysis Consortium (DPAC, \url{https://www.cosmos.esa.int/web/gaia/dpac/consortium}). Funding for the DPAC has been provided by national institutions, particularly those participating in the {\it Gaia} Multilateral Agreement.

% sponsors
TCG is supported by the European Union's Horizon 2020 research and innovation program under grant agreement No 101004110.

%% The Appendices part is started with the command \appendix;
%% appendix sections are then done as normal sections
\appendix

%%\section{Appendix title 1}
%% \label{}

%%\section{Appendix title 2}
%% \label{}

%% If you have bibdatabase file and want bibtex to generate the
%% bibitems, please use
%%
\bibliographystyle{elsarticle-harv} 
\bibliography{biblio}

%% else use the following coding to input the bibitems directly in the
%% TeX file.

%%\begin{thebibliography}{00}

%% \bibitem[Author(year)]{label}
%% For example:

%% \bibitem[Aladro et al.(2015)]{Aladro15} Aladro, R., Martín, S., Riquelme, D., et al. 2015, \aas, 579, A101

%%\end{thebibliography}

\end{document}